\documentstyle[12pt]{article}

\catcode`\@=11
\def\marginnote#1{}
\newcount\hour
\newcount\minute
\newtoks\amorpm
\hour=\time\divide\hour by60
\minute=\time{\multiply\hour by60 \global\advance\minute
by-\hour}\edef\standardtime{{\ifnum\hour<12
\global\amorpm={am}%
        \else\global\amorpm={pm}\advance\hour by-12 \fi
        \ifnum\hour=0 \hour=12 \fi
        \number\hour:\ifnum\minute<10
0\fi\number\minute\the\amorpm}}
\edef\militarytime{\number\hour:\ifnum\minute<10
0\fi\number\minute}

\def\draftlabel#1{{\@bsphack\if@filesw {\let\thepage\relax
   \xdef\@gtempa{\write\@auxout{\string
      \newlabel{#1}{{\@currentlabel}{\thepage}}}}}\@gtempa
   \if@nobreak \ifvmode\nobreak\fi\fi\fi\@esphack}
        \gdef\@eqnlabel{#1}}
\def\@eqnlabel{}
\def\@vacuum{}
\def\draftmarginnote#1{\marginpar{\raggedright\scriptsize\tt#1}}
\def\draft{\oddsidemargin -.5truein
        \def\@oddfoot{\sl preliminary draft \hfil
        \rm\thepage\hfil\sl\today\quad\militarytime}
        \let\@evenfoot\@oddfoot \overfullrule 3pt
        \let\label=\draftlabel
        \let\marginnote=\draftmarginnote

\def\@eqnnum{(\theequation)\rlap{\kern\marginparsep\tt\@eqnlabel}%
\global\let\@eqnlabel\@vacuum}  }

\def\numberbysection{\@addtoreset{equation}{section}
        \def\theequation{\thesection.\arabic{equation}}}

\def\underline#1{\relax\ifmmode\@@underline#1\else
 $\@@underline{\hbox{#1}}$\relax\fi}

\catcode`@=12
\relax

\numberbysection

\advance \topmargin by -\headheight
\advance \topmargin by -\headsep

\evensidemargin \oddsidemargin
\def\sd{D\!\!\!/}
\def\sp{\partial\!\!\!/}

\def\sb{b\!\!\!/}
\def\ss{s\!\!\!/}

\def\br{\begin{eqnarray}}
\def\er{\end{eqnarray}}
\def\be{\begin{equation}}
\def\ee{\end{equation}}

\def\({\left(}
\def\){\right)}

\relax

\def\a{\alpha}

\def\b{\beta}

\def\d{\delta}
\def\D{\Delta}

\def\g{\gamma}

\def\l{\lambda}
\def\L{\Lambda}

\def\pa{\partial}

\def\s{\sigma}
\def\S{\Sigma}

\def\tp0{\Theta_{+}^{(0)}}
\def\tm0{\Theta_{-}^{(0)}}

\def\f#1#2#3 {f^{#1#2}_{#3}}

\def\win1{{\sf w_{1+\infty}}}

\def\Win1{{\sf W_{1+\infty}}}

\def\rlx{\relax\leavevmode}
\def\inbar{\vrule height1.5ex width.4pt depth0pt}
\def\IZ{\rlx\hbox{\sf Z\kern-.4em Z}}
\def\IR{\rlx\hbox{\rm I\kern-.18em R}}
\def\IC{\rlx\hbox{\,$\inbar\kern-.3em{\rm C}$}}
\def\IN{\rlx\hbox{\rm I\kern-.18em N}}
\def\IO{\rlx\hbox{\,$\inbar\kern-.3em{\rm O}$}}
\def\IP{\rlx\hbox{\rm I\kern-.18em P}}
\def\IQ{\rlx\hbox{\,$\inbar\kern-.3em{\rm Q}$}}
\def\IF{\rlx\hbox{\rm I\kern-.18em F}}
\def\IG{\rlx\hbox{\,$\inbar\kern-.3em{\rm G}$}}
\def\IH{\rlx\hbox{\rm I\kern-.18em H}}
\def\II{\rlx\hbox{\rm I\kern-.18em I}}
\def\IK{\rlx\hbox{\rm I\kern-.18em K}}
\def\IL{\rlx\hbox{\rm I\kern-.18em L}}
\def\one{\hbox{{1}\kern-.25em\hbox{l}}}
\def\0#1{\relax\ifmmode\mathaccent"7017{#1}%
B        \else\accent23#1\relax\fi}

\def\EPJC#1#2#3{{\sl Eur. Phys. J.} {\bf C#1} (#2) #3}
\def\JGP#1#2#3{{\sl J. Geom. Phys.} {\bf #1} (#2) #3}
                \def\JHEP#1#2#3{{\sl JHEP} {\bf#1} (#2) #3}
                
                \def\NPB#1#2#3{{\sl Nucl. Phys.} {\bf B#1} (#2) #3}
                
                \def\CMP#1#2#3{{\sl Commun. Math. Phys.} {\bf #1} (#2) #3}
                \def\PRD#1#2#3{{\sl Phys. Rev.} {\bf D#1} (#2) #3}

                \def\PLB#1#2#3{{\sl Phys. Lett.} {\bf #1B} (#2) #3}
                \def\JMP#1#2#3{{\sl J. Math. Phys.} {\bf #1} (#2) #3}
                
                \def\PTP#1#2#3{{\sl Prog. Theor. Phys.} {\bf #1} (#2) #3}
                
                \def\AoP#1#2#3{{\sl Annals Phys.} {\bf #1} (#2) #3}

                \def\IJMPA#1#2#3{{\sl Int. J. Mod. Phys.} {\bf A#1} (#2) #3}

                \def\MPLA#1#2#3{{\sl Mod. Phys. Lett.} {\bf A#1} (#2) #3}

                \def\a{\alpha}
                \def\b{\beta}

                \def\d{\delta}
                \def\D{\Delta}
                
                \def\g{\gamma}

                \def\/{\frac}

                \def\l{\lambda}
                \def\L{\Lambda}

                \def\pa{\partial}

                \def\s{\sigma}

                \def\({\Big(}
                \def\){\Big)}
                \def\[{\Big[}
                \def\]{\Big]}

                \def\rlx{\relax\leavevmode}
                \def\inbar{\vrule height1.5ex width.4pt depth0pt}
                \def\IZ{\rlx\hbox{\sf Z\kern-.4em Z}}
                \def\IR{\rlx\hbox{\rm I\kern-.18em R}}
                \def\IC{\rlx\hbox{\,$\inbar\kern-.3em{\rm C}$}}
                \def\IN{\rlx\hbox{\rm I\kern-.18em N}}
                \def\IO{\rlx\hbox{\,$\inbar\kern-.3em{\rm O}$}}
                \def\IP{\rlx\hbox{\rm I\kern-.18em P}}
                \def\IQ{\rlx\hbox{\,$\inbar\kern-.3em{\rm Q}$}}
                \def\IF{\rlx\hbox{\rm I\kern-.18em F}}
                \def\IG{\rlx\hbox{\,$\inbar\kern-.3em{\rm G}$}}
                \def\IH{\rlx\hbox{\rm I\kern-.18em H}}
                \def\II{\rlx\hbox{\rm I\kern-.18em I}}
                \def\IK{\rlx\hbox{\rm I\kern-.18em K}}
                \def\IL{\rlx\hbox{\rm I\kern-.18em L}}
                \def\one{\hbox{{1}\kern-.25em\hbox{l}}}
                \def\0#1{\relax\ifmmode\mathaccent"7017{#1}%
                B        \else\accent23#1\relax\fi}
                
\begin{document}
\begin{titlepage}

                \begin{center}

                  {\large\bf Bosonized noncommutative bi-fundamental fermion and S-duality}

                \end{center}

\vspace{.5 cm}

                \begin{center}

                Harold Blas\\

                \vspace{.6 cm}

               Departamento de Matem\'atica - ICET\\
Universidade Federal de Mato Grosso\\
 Av. Fernando Correa, s/n, Coxip\'o \\
78060-900, Cuiab\'a - MT - Brazil\\

  \end{center}

                \begin{abstract}

                \vspace{.4 cm}

                 We perform the path-integral bosonization of the recently proposed noncommutative
massive Thirring model (NCMT$_{1}$) [JHEP0503(2005)037]. This
model presents two types of current-current interaction terms
related to the bi-fundamental representation of the group $U(1)$.
Firstly, we address the bosonization of a bi-fundamental free
Dirac fermion defined on a noncommutative (NC) Euclidean plane
$\IR_{\theta}^{2}$. In this case we show that the fermion system
is dual to two copies of the NC Wess-Zumino-Novikov-Witten model.
Next, we apply the bosonization prescription to the NCMT$_{1}$
model living on $\IR_{\theta}^{2}$ and show that this model is
equivalent to two-copies of the WZNW model and  a two-field
potential defined for scalar fields corresponding to the global
$U(1)\times U(1)$ symmetry plus additional bosonized terms for the
four fermion interactions. The bosonic sector resembles to the one
proposed by Lechtenfeld et al. [Nucl. Phys. B705(2005)477] as the
noncommutative sine-Gordon for a {\sl pair} of scalar fields. The
bosonic and fermionic couplings are related by a strong-weak
duality. We show that the couplings of the both sectors for some
representations satisfy similar relationships up to relevant
re-scalings, thus the NC bi-fundamental couplings are two times
the corresponding ones of the NC fundamental (anti-fundamental)
and eight times the couplings of the ordinary massive Thirring and
sine-Gordon models.
\end{abstract}

                \end{titlepage}

                \section{Introduction}
\label{intro}

Quantum field theories on non-commutative (NC) space-times are
receiving considerable attention in recent years in connection to
string and M theories (see, Refs. \cite{seiberg}). In the case of
open string with $N=2$ world-sheet super-symmetries and for target
space filled by $n$ coincident $D3-$branes the low-energy NC model
corresponds to NC self-dual $U(n)$ Yang Mills theory (NC SDYM)
\cite{Lechtenfeld1}. This theory is integrable classically
\cite{Takasaki} and possesses a factorized S-matrix in the quantum
version \cite{Lechtenfeld1}. It is a remarkable fact that almost
all NC integrable models in less than four dimensions can be
obtained by dimensional reduction of four dimensional NC SDYM (see
e.g. \cite{lechtenfeld} and references therein).

Some non-commutative versions of the sine-Gordon (NCSG$_{1\,,2}$)
\cite{lechtenfeld}-\cite{jhep2} and corresponding massive Thirring
models (NCMT$_{1,\,2}$) \cite{jhep2} have been proposed in the
literature. Their relevant NC equations of motion have the general
property of reproducing the ordinary models in the commutative
limit. The Grisaru-Penati version NCSG$_{2}$ \cite{grisaru1,
grisaru2} introduces a constraint which is non-trivial only in the
non-commutative case. However, at the quantum level this model
gives rise to particle production as was discovered by evaluating
tree-level scattering amplitudes \cite{grisaru2}. Moreover,
introducing an extra field, Lechtenfeld et al. \cite{lechtenfeld}
proposed a novel NCSG$_{1}$ model which presents a factorizable
and causal S-matrix at tree level computations. It seems to be
that requiring an infinite number of classical conserved
quantities does not uniquely reproduce the Moyal deformation of
two-dimensional integrable field theories. The models NCSG$_{1,
2}$ were recently derived from the NC self-dual $U(2)$ Yang Mills
theory through dimensional and algebraic reduction processes
\cite{grisaru2, lechtenfeld}. Moreover, these scalar field models
(NCSG$_{1, 2}$) were shown to be related to spinorial models on
the classical level, possessing soliton solutions in both sectors
and a strong-weak mapping of their relevant coupling constants.
These models, respectively named as
 NCMT$_{1,\,2}$, are Moyal extensions of the usual massive Thirring
model (MT) defined for a bi-fundamental($bf$) fermion with two
types of current-current interaction terms related to the group
$U(1)\times U(1)$ \cite{jhep2}. The NCMT$_{2}$ model requires two
copies of the NCMT$_{1}$ theory, this last model being just the NC
extension of the massive Thirring for a bi-fundamental fermion.

On the other hand, the equivalence of the ordinary sine-Gordon and
massive Thirring models is a two-dimensional example of having
different field representations of the same theory. The massive
Thirring model (MT) \br {\cal L}_{MT}&=& i\bar{\psi} \g^{\mu}
\pa_{\mu}  \psi +m \bar{\psi}  \psi -\frac{\l_{MT}}{2} j^{\mu}
j_{\mu} \label{mt0}, \er is equivalent to the sine-Gordon model
(SG) \cite{coleman} \br {\cal L}_{SG}&=& \frac{1}{2}
\pa_{\mu}\phi\pa^{\mu}\phi + \frac{\a_{0}}{\b^2_{SG}} \(\mbox{cos}
\,\b_{SG}\, \phi -1\)
 \label{sg0}. \er

This duality is expressed in terms of the well-known abelian
bosonization rules. The two field representations (\ref{mt0}) and
(\ref{sg0}) describing the same physics are useful in different
coupling regimes which are related by \br\label{dual0}
\frac{4\pi}{\b^2_{SG}}= 1+ \frac{\l_{MT}}{\pi}.\er

The particle/soliton correspondence in the context of this duality
 has been uncovered through the master Lagrangian and
  symplectic quantization approaches \cite{nucl2} of the higher
grading $sl(2)$ affine Toda model coupled to matter (Dirac) field
\cite{matter}. A generalization for any (untwisted) affine Lie
algebra has been provided in \cite{jhep1}.

Here we consider the problem of extending this duality to the NC
spacetime using the path-integral formalism. This program,
initiated in \cite{nunez}, considers the abelian bosonization of a
free fermion in two noncommuting dimensions \cite{moreno1,
moreno2} using path-integral techniques developed in refs.
\cite{Naon}-\cite{Guillou}. We point out that in these NC
treatments the Dirac fermion was assumed to be in the
 anti-fundamental($\bar{f}$) representation of the $U(1)$ group. Next, we
summarize the main features that emerged from this bosonization
procedure. The free fermion action is bosonized to a NC $U(1)$
WZW-action in which the WZ term in the action gives a non-trivial
contribution due to the non-commutativity of spacetime. The
procedure resembles closely the conventional non-abelian
bosonization \cite{Witten1} with the bosonization rules being
similar both in the non-abelian and NC cases. The massive Thirring
model with anti-fundamental fermion is dual to a WZW model plus a
NC cosine potential and an additional quartic term emerging from
the NC bosonization of the current-current interaction. In this
way, except for the quartic term, a copy of the Grisaru-Penati
model emerges in the bosonic sector \cite{nunez}. Recall that the
Grisaru-Penati model \cite{grisaru1, grisaru2, jhep2} is defined
by the sum of a WZNW action for
$e^{i\b_{\bar{f}}\,\phi}_{\star}\in U(1)_{C}$ and a cosine
potential $(e^{i\b_{\bar{f}}
\phi}_{\star}+e^{-i\b_{\bar{f}}\phi}_{\star})$, plus some copies
of these terms  written for
$e^{i\b_{\bar{f}}\,\phi^{\dagger}}_{\star}$,\, $\b_{\bar{f}}$
being the bosonic coupling constant. The same type of duality
relationship, eq. (\ref{dual0}), emerges up to couplings
re-scalings \cite{nunez} \br\label{dual01}
\frac{16\pi}{\b^2_{\bar{f}}}= 1+ \frac{\l_{\bar{f}}}{4\pi},\er
 thus realizing an example of S-duality in NC spacetime. The S-duality in the context of NC geometry has
 also been discussed in Refs. \cite{Ganor}.

The Grisaru-Penati model, in spite of integrability in the sense
that it possesses an infinite number of conserved quantities,
features an acausal S-matrix and particle production takes place.
It was recently shown, on the classical level, that this model
corresponds to the Moyal product extension of two copies of the
ordinary massive Thirring model written for a bi-fundamental
fermion \cite{jhep2}. The main features observed in this classical
correspondence, apart from the doubling in the number of fermion
fields, is that, even though the strong-weak duality in the
soliton sector is preserved, only the unitary subgroup of the
$U(1)_{C} \times U(1)_{C}$ symmetry of the bosonic model is
realized in the spinorial model.

In two-dimensional Minkowski spacetime, a Moyal deformation
necessarily implies the noncommutativity of the time coordinate
and the causality and unitarity properties of the theory can be
spoiled \cite{Gomis}. However, in an exactly solvable quantum
field theory this feature can be improved as has been observed in
the NC Lee model, where it was shown that the model is free from
the $IR/UV$ mixing in both the space-space and space-time
noncommutative cases. Due to the absence of this type of mixing
one can expect that the theory is unitary \cite{Chu1}.

In this context, it would be interesting to understand what
actually determines the systems to be integrable (in the sense of
possessing a factorizable and causal S-matrix) and dual to each
other. From the considerations summarized above on NC extensions
of SG model, the correspondence NCSG$_{1}$ $\rightarrow$
NCMT$_{1}$ is promising and deserves a further investigation. The
main features of the NCSG$_{1}$ model are the
 presence of multi-soliton solutions, the factorizable and causal
 S-matrix (verified at tree level) despite of the space-time noncommutativity \cite{lechtenfeld}.
 Moreover,  these models originate directly through  reduction processes starting from the NC WZNW
type action for the $GL(2)$ affine Toda field coupled to the
higher grading matter (Dirac) fields (NCATM$_{1}$), such that the
$U(1) \times U(1)$ symmetry is relevant, both in the construction
of the NCSG$_{1}$ model and in the star-localized Noether
procedure to construct the $U(1)\times U(1)$ currents in the
NCMT$_{1}$ sector \cite{jhep2}. Then, it is interesting to perform
the bosonization of the massive Thirring theory (NCMT$_{1}$) as
defined in
 \cite{jhep2} and study the corresponding bosonic theory properties; e.g. to answer the question of what features
 it shares with the classical NCSG$_{1}$ theory.

The paper is organized as follows. In the next section we perform
the NC bosonization of the bi-fundamental free fermion. In section
\ref{mth} we bosonize the NCMT$_{1}$ theory making use of the
bosonization rules for the NC free fermion. In subsection
\ref{mthint} we bosonize the current-current interaction terms and
identify the canonical bosonic physical field, related to the
global $U(1)$ charge, coming from the kinetic terms and from the
boson quartic interactions. In subsection \ref{mthmass} we
bosonize the mass term by axial symmetry consideration alone and
get a two-field potential. Next, considering the physical field
 and the relevant coupling
constants we establish the S-duality for the NC field theories
under consideration. Moreover, we record the bosonized model in
matrix form which is of the type proposed by Lechtenfeld et al.
(NCSG$_{1}$) except for the quartic contributions coming from the
 current-current interactions. Finally, the conclusions and
discussions are presented in section \ref{conclusion}.

\section{The bi-fundamental fermion and NC bosonization}
\label{freenc}

A conventional system of $N$ free massless Dirac fermions is
equivalent to a bosonic theory governed by the WZW model for a
bosonic field $g \in SU(N)$ and a real bosonic field $\phi$ with a
free scalar action \cite{Witten1}. On the other hand, the massless
free fermion in the fundamental representation or anti-fundamental
representation \cite{moreno1, moreno2, nunez} of the abelian
$U(1)$ group on $\IR_{\theta}^{2}$ is equivalent to a
noncommutative version of the WZW model \cite{moreno1, Furuta}. In
this section we derive the bosonization rules for the
bi-fundamental free fermion action defined on $\IR_{\theta}^{2}$.
We follow the path integral approach developed in \cite{Burgess2,
Guillou, nunez}.

We consider the bi-fundamental $U(1)\times U(1)$ free fermion
theory \br \label{free0} S_{bf}=\int d^2 x\,i\,
\bar{\psi}^{j_{2}}_{\, \,\,i_{1}} \d_{\,\,i_{2}}^{i_{1}}
\d_{\,\,j_{2}}^{j_{1}}\,\star \sp \psi^{i_{2}}_{\,\,j_{1}}, \er
where the group indices $i$ corresponds to the first $U(1)$ and
the indices $j$ to the second $U(1)$. We will see that this theory
is equivalent to a bosonic model by showing that the correlation
functions constructed from the both theories are equal. So, let us
write the generating functional for the correlations as
 \br\label{partifree1}
Z[s_{1},\, s_{2}]= \int {\cal D}\bar{\psi} \,{\cal D}\psi \,
\mbox{exp}\{-\int d^2 x\, \bar{\psi}^{j_{1}}_{\,\,i_{1}}\star (i
\sd [s^{(1)},\, s^{(2)}] \psi )^{i_{1}}_{\,j_{1}}\} \er where
 \br \{D_{\mu}[s^{(1)},
s^{(2)}] \psi(x)\}^{i_{1}}_{\,\,j_{1}} \equiv
\[\pa_{\mu}\d_{\,\,i_{2}}^{i_{1}}
\d_{\,\,j_{1}}^{j_{2}} + i  \star \d^{i_{1}}_{\,\,i_{2}}\,
s_{\mu\,\,\,j_{1}}^{(1)j_{2}}- i \d^{j_{2}}_{\,\,j_{1}}\,
s^{(2)i_{1}}_{\mu\,\,\,i_{2}} \star
\] \psi(x)^{i_{2}}_{\,\,j_{2}} , \er with the notation $[\star s^{(1)}_{\mu}(x)]\psi(x)
\equiv \psi(x) \star s^{(1)}_{\mu}(x)$\, and
$[s^{(2)}_{\mu}(x)\star] \psi(x) \equiv s^{(2)}_{\mu}(x) \star
\psi(x)$.

Notice that in the above generating functional we have considered
 two types of external sources, i.e. $s_{\,\mu}^{(1)}$ and
$s_{\,\mu}^{(2)}$, such that differentiations with respect to each
source provide correlation functions between the relevant currents
\br\label{current1} j^{(1)\,\mu} &\equiv& \bar{\psi}\star \g^{\mu}
\psi;\\\label{current2} j^{(2)\,\mu}&\equiv& \psi_{\b} (
\g^{\mu})_{\a\b} \star \bar{\psi}_{\a},\er respectively.

The currents $j^{(1)\, \mu}$ and $j^{(2)\, \mu}$ as defined above
differ only by a sign in the commutative limit, whereas on NC
Euclidean space they are different \cite{gracia-bondia}. Next, we
briefly describe the steps followed to obtain the currents
(\ref{current1})-(\ref{current2}) associated to a bi-fundamental
fermion system. In order to obtain the currents by the Noether
procedure we make the global transformation localized, as
discussed in \cite{liao} the star-localized procedure is not
unique in the NC case. In fact, the most general $U(1)\mbox{x}
U(1)$ symmetry in NC space given for a charged bi-fundamental
 field is provided by the transformation rules \cite{liao, terashima}
 \br \label{u1u1nc}
\psi \rightarrow u_{2}(x) \star \psi \star
u_{1}^{-1}(x);\,\,\,\,\bar{\psi} \rightarrow u_{1}(x) \star
\bar{\psi} \star u_{2}^{-1}(x), \,\,\,\,u_{1}(x)=e_{\star}^{-i
\a_{1}(x)},\,\, u_{2}(x)=e_{\star}^{i \a_{2}(x)},\er where
$u_{1,2}(x)$ are independent starred exponentials with
$\a_{1,2}(x)=\mbox{real functions}$. The two currents above
(\ref{current1})-(\ref{current2}) correspond to the symmetries
$u_{1}$ and $u_{2}$, respectively.

Since the charge is associated to global transformation of the
charged field for which there is no difference between the
ordinary and non-commutative product, one can conclude that the
currents $j^{(1)\,\mu}$\, and \,$j^{(2)\,\mu}$  share the same
charge. In fact, for global $u_{1, 2}$ only the product \br
\label{global}u_{{\sl global}}= u_{2} u^{-1}_{1}\er is relevant.

In the following we use the approach developed in \cite{moreno2,
nunez} conveniently adapted to our problem. The measure in
(\ref{partifree1}) is invariant under the local transformations of
the fermion fields (\ref{u1u1nc}) and consequently the generating
functional (\ref{partifree1}) is gauge invariant under separate
gauge transformations acting on $s^{(1)}$ and $s^{(2)}$, namely
 \br
 \label{identity}
Z[s^{(1)},\, s^{(2)}]\,=\, Z[s^{(1)\,U_{1}},\, s^{(2)\,U_{2}}]
 \er
provided that the sources transform according to \br
\label{gauget} s^{(n)}_{\mu} \rightarrow s_{\mu}^{(n)\,U_{n}} =
U_{n}\star s_{\mu}^{(n)}\star U_{n}^{-1} + U_{n} \pa_{\mu}
U_{n}^{-1},\,\,\,\,\,n=1,2.\er

Then from the identity (\ref{identity}) one gets \br \nonumber
Z[s^{(1)},\, s^{(2)}] &=& \int {\cal D}\bar{\psi} \,{\cal D}\psi
\, {\cal D}U_{1}\,{\cal D} U_{2}\,  \mbox{exp}\{-\int d^2 x\,
\bar{\psi}( i\star \sp -\star
\ss^{(1)\,U_{1}}+\ss^{(2)\,U_{2}}\star)\psi\}
\\&=& \int {\cal D} U_{1}\,{\cal D} U_{2}\,  \mbox{det}(
i\,  \sp-\star\,\ss^{(1)\,U_{1}}+ \ss^{(2)\,U_{2}} \star
),\label{integrated}\er where the fermions have been integrated
out. Notice that the integration over the $U_{n}$'s just amounts
to a change of the overall normalization.

It is useful to introduce an explicit representation of the group
$g \in U(1)\times U(1)$ in the form \br \label{u1u1} g =
\left(\begin{array}{cr}
e^{i\L_{2}}_{\star} &  0 \\
0 & e^{-i\L_{1}}_{\star}
\end{array} \right)\,\equiv\, g_{2}\star  g_{1}= g_{1}\star g_{2},\,\,\,\,
g_{2} = \left(\begin{array}{cr}
 e^{i\L_{2}}_{\star}&  0 \\
0 & 1
\end{array} \right),\,\,\,\, g_{1} = \left(\begin{array}{cr}
1 &  0 \\
0 &  e^{-i\L_{1}}_{\star}
\end{array} \right),
\er where $\L_{1,2}$ are real fields.

 In this $2\times 2$ representation the matrix Lie
algebra valued fields $\ss^{(n)}\,(n=1,2)$  corresponding to each
subalgebra respectively has just one entry different from zero,
respectively. Then the matrix of interest becomes \br \nonumber(
i\star \sp -\star \ss^{(1)\,U_{1}}+\ss^{(2)\,U_{2}}\star)_{2\times
2} &=& \left(\begin{array}{cc}
 i\sp+\ss^{(2)\,U_{2}}\star &  0 \\
0 &   i\sp-\star\,\ss^{(1)\,U_{1}}
\end{array} \right)\\&=&\left(\begin{array}{cc}
 i\sp+ \ss^{(2)\,U_{2}} \star &  0 \\
0 & 1
\end{array} \right)\left(\begin{array}{cc}
1   &  0 \\
0 & i \sp -\star \ss^{(1)\,U_{1}}
\end{array} \right),\label{matprod}
\er where in the right hand sides of (\ref{matprod}) the fields
$s^{(n)\,U_{n}}\,(n=1,2)$ denote just functions instead of Lie
algebra valued fields.

 Therefore the generator functional factorizes \br \label{factorized1} Z[s^{(1)},\,
s^{(2)}] &=& \int {\cal D}U_{1}\,{\cal D} U_{2}\,\,\mbox{det}(
i\sp-\star\,\ss^{(1)\,U_{1}})
\,\,\mbox{det}(i\sp+\ss^{(2)\,U_{2}}\star),\er where the Lie
algebra valued character has been restored for the fields
$\ss^{(n)\,U_{n}}$.

 This factorization is expected since $Z[s^{(1)},\,
s^{(2)}]$ is gauge invariant under the separate gauge
transformations (\ref{gauget}) acting on $s^{(1)}$ and $s^{(2)}$,
respectively

 Let us introduce the connections through
\br b_{\mu}^{(n)} = s_{\mu}^{(n)\, U_{n}},\er such that the field
strengths of $b_{\mu}^{(n)}$  and $s_{\mu}^{(n)}$ satisfy \br
F_{\mu\nu}^{(n)}[b^{(n)}] =U_{n} \star
F_{\mu\nu}^{(n)}[s^{(n)}]\star U_{n}^{-1},\,\,\,\,n=1,2.\er

We proceed by changing the $U_{n}$ integrations
 to integrations over the new connections $b^{(n)}$. In order to achieve that we
 shall use the identity \cite{Guillou}
 \br
 \label{stren}
 \int {\cal D}b_{\mu}^{(n)} {\cal P}[b_{\mu}^{(n)}] \d[\epsilon^{\mu \nu} (F_{\mu\nu}^{(n)}[b^{(n)}]-
F_{\mu\nu}^{(n)}[s^{(n)}])] =\int {\cal D}U_{n}{\cal
P}[s_{\mu}^{(n)\, U_{n}}], \,\,\,\,\mbox{for}\,\,n=1,2.
 \er
where ${\cal P}$ is a gauge invariant function.

Making use of the identities (\ref{stren})  the generating
functional (\ref{factorized1}) can be written as \br Z[s^{(1)},\,
s^{(2)}] = \int {\cal D}b_{\mu}^{(n)} \D_{n}\,\,
\d(b^{(n)}_{+}-s^{(n)}_{+})\,\, \d[\epsilon^{\mu \nu}
(F_{\mu\nu}^{(n)}[b^{(n)}]- F_{\mu\nu}^{(n)}[s^{(n)}])]\nonumber
\\
\times \mbox{det}( i\sp-\star\,\sb^{(1)})\,\mbox{det}(
i\sp+\sb^{(2)}\star ),\label{gfix1}\er where we have chosen the
gauge conditions $b_{+}^{(n)}=s_{+}^{(n)},\,\,n=1,2$; with the
$\D_{n}$'s being the relevant Faddeev-Popov determinants.

We define the Lie algebra valued fields $\hat{a}_{n}$ and use them
in order to reinforce the delta function conditions, i.e. \br
\nonumber Z[s^{(n)}] &=& \int {\cal D}\hat{a}_{n} {\cal
D}b_{\mu}^{(n)}\, \D_{n} \,\,\d(b^{(n)}_{+}-s^{(n)}_{+}) \,\,
\mbox{det}\(i\sp-\star\,\sb^{(1)} \) \mbox{det}\(i\sp+\sb^{(2)} \star \)\\
&& \times\,\, \Pi_{n} \mbox{exp} \({\cal
A}_{n}\)\,\label{gfix11}\er where ${\cal A}_{n} \equiv -
\frac{C_{n}}{8\pi}\mbox{tr}\, \int d^2x \,\hat{a}_{n}\,
\d[\epsilon^{\mu \nu} (F_{\mu\nu}^{(n)}[b^{(n)}]-
F_{\mu\nu}^{(n)}[s^{(n)}])]$ , with the $C_{n}$'s being some
constants to be conveniently chosen below. In two dimensions one
can parameterize a gauge boson as a pure gauge field, so the
source fields $s_{\mu}^{(n)}$ and $b_{\mu}^{(n)}$ can be written
in terms of group-valued variables \br \label{par1}s_{+}^{(n)}&=&i
\widetilde{s}_{n}^{\,-1} \pa_{+} \widetilde{s}_{n}\\\label{par2}
s_{-}^{(n)}&=&i s_{n} \pa_{-}s_{n}^{-1}\\\label{par3}
b_{+}^{(n)}&=&i (\widetilde{b}_{n}\widetilde{s}_{n} )^{-1} \pa_{+}
( \widetilde{b}_{n}\widetilde{s}_{n})\\\label{par4}
b_{-}^{(n)}&=&i (s_{n}b_{n}) \pa_{-}(s_{n}b_{n})^{-1}
 \er

The $U(1)\times U(1)$ group valued fields $s_{n},
\widetilde{s}_{n}, b_{n}, \widetilde{b}_{n}$ are in the same
representation as the $g_{n}$'s \,$(n=1,2)$ in
 (\ref{u1u1}).

Notice that the determinant depending on $\sb_{2}$ is associated
to the fundamental fermion representation and the one with
$\sb_{1}$ resembles to that of the anti-fundamental. The
fundamental and anti-fundamental determinants coincide, as
discussed in \cite{moreno2}. The exact effective action in each
case is obtained by integrating the chiral anomaly \cite{moreno1,
moreno2}. On the other hand, in two dimensions the bi-fundamental
Dirac fermion carries no mixed anomaly \cite{nakajima, Bonora}
(for chiral matter see \cite{Martin}). Therefore, one can make the
affective action computation separately for each $U(1)$ sector.
Taking into account (\ref{factorized1}) the fermion determinant
for the parameterizations (\ref{par1})-(\ref{par4}) can be written
as \cite{moreno1, moreno2} \br \mbox{det}(
i\sp-\star\,\sb^{(1)}+\sb^{(2)} \star ) = \mbox{exp}
(W[\widetilde{b}_{1}\star \widetilde{s}_{1} \star s_{1}\star
b_{1}])\,\, \mbox{exp} (W[\widetilde{b}_{2}\star \widetilde{s}_{2}
\star s_{2} \star b_{2}]), \er where \br\label{wznw}
W[g]=\frac{1}{8\pi} \int_{\S} d^2x \mbox{Tr} (\pa_{\mu}g \star
\pa_{\mu}g^{-1}) -\frac{i}{12\pi} \int_{B} d^3 x \epsilon^{ijk}
\mbox{Tr}(g^{-1}\star \pa_{i}g \star g^{-1}\star \pa_{j}g \star
g^{-1}\star \pa_{i}g),\er is the NC extension of the WZNW model
\cite{moreno1, Furuta}. The manifold $\S$, parameterized by $(x^0,
x^1)$, is the boundary of the three-dimensional manifold $B$. The
extra dimension $x^2$ is taken to be commutative.

The Jacobians for the change of variables from $b_{\mu}^{(n)}$ to
 $b_{n},\,\widetilde{b}_{n}$ become \cite{nunez}
 \br {\cal D}b_{+}^{(n)}{\cal D}b_{-}^{(n)}&=& \mbox{det}_{\star} D_{+}[\widetilde{b}_{n} \star
\widetilde{s}_{n}]\, \mbox{det}_{\star} D_{-}[s\star b] {\cal
D}\widetilde{b}_{n}\,{\cal D}b_{n}\,\nonumber
\\&=& {\cal D}\widetilde{b}_{n}\,{\cal D}b_{n}\, \mbox{exp}
(\kappa_{n} W[\widetilde{b}_{n}\star \widetilde{s}_{n}\star
s_{n}\star b_{n}]),\, \,\,\,\,\,n=1,2; \er where $\kappa_{n}$ is
the parameter depending on the representation of the covariant
derivative involved in the change of variables, and since we shall
not use it here we leave it as an unspecified parameter.

The delta functional in (\ref{gfix11}), for these change of
variables,  can be written as \br
\d(b^{(n)}_{+}-s^{(n)}_{+})=\frac{1}{\mbox{det}_{\star}D_{+}[s^{(n)}_{+}]}
\d(\widetilde{b}^{(n)}-I).\er

Then, from the results above, (\ref{gfix11}) can be written as \br
 Z[s^{(n)}] &=& \int {\cal D}\hat{a}_{n} {\cal
D}b_{n}\,\,\, \Pi_{n} \mbox{exp}({\cal B}_{n})\,\, \,\mbox{exp}
\((1+\kappa_{n})
 W[\widetilde{s}_{n}\star s_{n}\star b_{n}]\),\label{group1}\er
 where
 \br
{\cal B}_{n} =i \frac{C_{n}}{4\pi} \mbox{tr}\, \int d^2x
(D_{+}[\widetilde{s}_{n}]\star \hat{a}_{n}) s_{n}\star b_{n}\star
(\pa_{-}b_{n}^{-1}) \star s_{n}^{-1},
 \er
and the covariant derivative is in the adjoint representation of
the corresponding $U(1)$ group.

In the present approach to bosonization one introduces two
group-valued fields, each one related to the corresponding $U(1)$
group, say $g_{1}$ and $g_{2}$, which will play the role of the
boson fields equivalent to the original bi-fundamental fermionic
model and will be invariant under gauge transformations. So, let
us introduce a change of variables from the algebra-valued fields
$\hat{a}_{n}$ to the group-valued fields $a_{n}$ as \br
D_{+}[\widetilde{s}_{n}]\star \hat{a}_{n}= \widetilde{s}_{n}^{\,
-1}\star(a^{-1}_{n} \star\pa_{+} a_{n}\star \widetilde{s}).\er

The Jacobian of these transformations are \cite{Burgess2, nunez}
\br {\cal D} \hat{a}_{n} =J_{L}^{(n)} {\cal
D}a_{n},\,\,\,\,\,J_{L}^{(n)}=\mbox{exp} (\kappa_{n} W[a_{n}\star
\widetilde{s}_{n}\star s_{n}]-\kappa_{n} W[\widetilde{s}_{n}\star
s_{n}]),\,\,\,\,\,n=1,2.\er

Therefore, the generating functional becomes \br
 Z[s^{(n)}] &=& \int {\cal D}a_{n} {\cal
D}b_{n}\, \Pi_{n} \,\mbox{exp} \((1+\kappa_{n})
 W[\widetilde{s}_{n}\star s_{n}\star b_{n}]+ \kappa_{n}  W[a_{n}\star \widetilde{s}_{n}\star s_{n}]- \kappa_{n}
 W[\widetilde{s}_{n}\star s_{n}]-\nonumber \\&&
 \frac{C_{n}}{4\pi} \mbox{tr} \int d^2 x \,
\widetilde{s}_{n}^{\,-1}\star (a^{-1}_{n}\star \pa_{+} a_{n})\star
\widetilde{s}_{n}\star s_{n}\star (b_{n}\star \pa_{-} b_{n}^{-1})
\star s_{n}^{-1} \) \label{group2}.\er

Choosing the up to now arbitrary constants $C_{n}$ to be
$C_{n}\equiv (1 + \kappa_{n})$ and making use of the
Polyakov-Wiegmnann identity \br \label{PW} W[g_{2}\star g_{1}]=
W[g_{2}]+ W[g_{1}]-\frac{1}{4\pi} \int_{\S} d^2x \mbox{Tr}
(g_{2}^{-1}\star \pa_{+} g_{2} \star g_{1} \star \pa_{-}
g_{1}^{-1} ),\er

one can write \br Z[s^{(n)}] = \int {\cal D}a_{n} {\cal D}b_{n}\,
\Pi_{n} \,\mbox{exp} \((1+\kappa_{n})
 W[a_{n}\widetilde{s}_{n}s_{n}b_{n}]
+ W[\widetilde{s}_{n}s_{n}]-W[a_{n}\widetilde{s}_{n}s_{n}] \)
\label{group3}\er

Since the integrations over the $b_{n}$'s can be trivially
factorized by making the changes $b_{n} \rightarrow
\hat{b}_{n}=a_{n}\widetilde{s}_{n}s_{n}b_{n}$, each one with
trivial Jacobian, one has \br Z[s^{(n)}] = \int {\cal D}a_{n}
\Pi_{n} \,\mbox{exp} \(
W[\widetilde{s}_{n}s_{n}]-W[a_{n}\widetilde{s}_{n}s_{n}]\)
\label{group31}.\er

The final transformations $a_{n}\widetilde{s}_{n}s_{n} \rightarrow
\widetilde{s}_{n}a_{n}s_{n}$ with trivial Jacobians; together with
the Polyakov-Wiegmann identity provide us the boson counterpart of
the bi-fundamental fermion model in (\ref{partifree1}) (as usual
rename $a_{n}$ as $g_{n}$ )\br Z[s^{(n)}_{\pm}] &=& \int \Pi_{n}
{\cal D}g_{n} \,\mbox{exp} \[ -W[g_{n}] + \frac{i}{4\pi} \mbox{tr}
\int d^2 x \, (s_{+}^{(n)}\star g_{n} \star \pa_{-} g_{n}^{-1} +
s_{-}^{(n)} \star g_{n}^{-1}\star \pa_{+} g_{n})+
 \nonumber\\
&& \frac{1}{4\pi} \mbox{tr} \int d^2 x \, (g_{n}^{-1}\star
s_{+}^{(n)}\star g_{n}\star s_{-}^{(n)}-s_{+}^{(n)}\star
s_{-}^{(n)})\],
 \label{group4}\er
where the matrices $g_{i}$ are in the representation (\ref{u1u1}).

Differentiating (\ref{group4}) with respect to any of the
components of the sources gives the correlation functions in the
relevant chirality sector \br \label{bosonimatrix}
j_{+}^{(n)}\rightarrow \frac{i}{4\pi} \mbox{tr}\, g_{n}^{-1}\star
\pa_{+} g_{n}, \,\,\,\, j_{-}^{(n)}\rightarrow \frac{i}{4\pi}
\mbox{tr}\, g_{n}\star  \pa_{-} g_{n}^{-1}, \,\,\,\,\,n=1,2;\er or
in terms of the field components
 \br\label{bosoni1}
 \bar{\psi}{\star}\g_{+} \psi & \rightarrow & \frac{i}{4\pi}
 e^{i\L_{1}}_{\star} \star
 \pa_{+} e^{-i\L_{1}}_{\star},\,\,\,\,\,\,\,\,\,\,\,\,\,\,\,\,\bar{\psi} {\star}\g_{-} \psi\,\,\,\,\,
 \rightarrow \,\,\,\,\,\,\,\frac{i}{4\pi} e^{-i\L_{1}}_{\star}
 \star
 \pa_{-} e^{i\L_{1}}_{\star};\\\label{bosoni2}
\psi_{\b} {\star}(\g_{+})_{\a \b} \bar{\psi}_{\a} & \rightarrow &
\frac{i}{4\pi} e^{-i\L_{2}}_{\star}
  \star \pa_{+} e^{i\L_{2}}_{\star},\,\,\,\psi_{\b} {\star}(\g_{-})_{\, \a\b} \bar{\psi}_{\a} \,\,\,\,
 \rightarrow\,\,\,\,\, \frac{i}{4\pi}
 e^{i\L_{2}}_{\star} \star
 \pa_{-} e^{-i\L_{2}}_{\star}.\er

Taking into account the representation (\ref{u1u1}) one can see
that the global free fermion symmetries $U(1)\times U(1)$ of type
(\ref{u1u1nc}) translate to the bosonized model (\ref{group4})  as
\br g_{2} \rightarrow U_{2}\, g_{2},\,\,\,\,g_{1} \rightarrow
U_{1}^{-1}\, g_{1},\,\,\,\,U_{2}=\mbox{diag}(u_{2},
1),\,\,\,\,U_{1}=\mbox{diag}(1, u_{1}). \label{globalbosmat}\er

Then, considering  the product $u_{2} u_{1}^{-1}$
 ($\a_{i}=$const.) as in (\ref{global}) for the global $U(1)$, the
transformations (\ref{globalbosmat}) written for the product of
field components becomes \br e^{i\L_{2}}_{\star}\star
e^{i\L_{1}}_{\star}
 \,\rightarrow\, u_{2} u^{-1}_{1}\, \, e^{i\L_{2}}_{\star}\star e^{i\L_{1}}_{\star} \label{pm0}.\er

We consider the bosonic theory (\ref{group4}) in terms of the
boson fields that appear naturally in the commutative limit
corresponding to the field responsible for the global $U(1)$
charge and a decoupled free field, respectively (similar arguments
have been used in \cite{liao} to study a  gauge theory of a doubly
 $U(1)$ gauged matter in NC spacetime). So, in the $\theta \rightarrow 0$
the transformation (\ref{pm0}) becomes \br
e^{i\L_{+}}\,\rightarrow\, u_{2} u^{-1}_{1}\, \,
e^{i\L_{+}},\,\,\,\,\L_{\pm} \equiv (\L_{2}\pm \L_{1})
\label{pm1}.\er

Then, the charge corresponding to the global $U(1)$ symmetry of
the NC free fermion should be related to the field $\L_{+}$ of the
bosonic sector. Once the physical fields are identified, one can
re-write the action of (\ref{group4}) in terms of them on NC
spacetime. Thus, writing the action of (\ref{group4}) in terms of
the fields $\L_{\pm}$ one can see that these fields do not
decouple from each other, whereas in commutative space they do.
Let us discuss this point in more detail using the
Polyakov-Wiegmnann identity (\ref{PW}).
 Since the last term in (\ref{PW}) vanishes for the representation
(\ref{u1u1}), then the sum $W[g_{1}]+ W[g_{2}]$ is equivalent to
$W[g_{2}\star g_{1}]$ which couples the fields $\L_{+}$ and
$\L_{-}$ in a non-trivial way. However, it can be seen that the
fields $\L_{-}$ and $\L_{+}$ decouple in the commutative limit of
the action (\ref{group4}). In this limit one has $W[g_{2}g_{1}]=
\int d^2x
\frac{1}{16\pi}[(\pa_{\mu}\L_{+})^2+(\pa_{\mu}\L_{-})^2]$. Since
the bosonized action of an ordinary free Dirac fermion in
Minkowski spacetime corresponds just to one real scalar field,
then one can set $\L_{1}=\L_{2}$ ($\L_{-}=0$) in the commutative
limit (however, see in the last paragraph of the next section a
discussion regarding the complex character of the fields $\L_{i}$
 in the bosonization process). Moreover, in this limit
the two currents differ only by a minus sign; in fact, by
comparing (\ref{bosoni1}) and (\ref{bosoni2}) one has
$j^{(1)}_{\mu}=-j^{(2)}_{\mu}$, as pointed out before. Regarding
the discussions above, an important point is that the symmetry
properties are simplest in terms of the fields $\L_{1}$ and
$\L_{2}$ and the physical interpretation becomes direct when
expressed in terms of the fields $\L_{\pm}$.

\section{The NC bi-fundamental massive Thirring
model} \label{mth}

The (Euclidean) Lagrangian of the non-commutative massive Thirring
model (NCMT$_{1}$) \footnote{Here we consider the fermionic analog
of the (bosonic) model NCMT$_{1}$ of Ref. \cite{jhep2}.} is

\br {\cal L}_{NCMT}&=&  i\bar{\psi} \g^{\mu} \pa_{\mu} \star \psi
+m \bar{\psi} \star \psi -\frac{\l_{bf}}{4} j^{(1)\,\mu} \star
j^{(1)}_{\mu} - \frac{\l_{bf}}{4} j^{(2)\,\mu}\star j^{(2)}_{\mu}
 \label{ncmt0}, \er
defined for Dirac fields and with the currents given in
(\ref{current1})-(\ref{current2}). Here, $\l_{bf}$ is the relevant
coupling constant and the group index contractions are being
assumed.

The action related to (\ref{ncmt0}) in terms of the spinor field
components becomes \br {\cal S}_{NCMT}&=&\int d^2x\,\,\[
 \nonumber
 i\widetilde{\psi}_{L}\pa_{+}\psi_{L} +
 i\widetilde{\psi}_{R}\pa_{-}\psi_{R}+ m \(\widetilde{\psi}_{R}\psi_{L}
+\widetilde{\psi}_{L}\psi_{R}\)-\\
&&\l_{bf}\(\widetilde{\psi}_{R}\star \psi_{R}\star
\widetilde{\psi}_{L}\star \psi_{L}+ \psi_{R}\star
\widetilde{\psi}_{R}\star \psi_{L}\star \widetilde{\psi}_{L}
\)\]\label{ncmt}, \er where again the group index notation has
been suppressed for convenience.

Notice that the NCMT model (\ref{ncmt0}) contains the non-standard
interaction term $j^{(2)}_{\mu}\star j^{(2)\,\mu} \sim
\psi_{R}\star\widetilde{\psi}_{R}\star\psi_{L}\star\widetilde{\psi}_{L}$,
which, to our knowledge, has not been considered previously in the
literature. Since the currents $j^{(1)\, \mu}$ and $j^{(2)\, \mu}$
differ only by a sign in the commutative limit, the Lagrangian
(\ref{ncmt0}) reduces to the usual MT model in this limit. The
bosonization process of the NC extension of the usual Thirring
interaction performed in \cite{moreno2, nunez} considers only
$j^{(1)}_{\mu}\star j^{(1)\,\mu}$. The inclusion of a second
$U(1)$ current will change various aspects of the bosonized model,
such as the potential, the duality relationship and the field
content, as we will show below.

The  $U(1)\times U(1)$ symmetries give the associated conservation
equations for the NCMT$_{1}$ model \br\label{curr11} \pa_{\mu} j^{(1)\,\mu}&=&0,\\
\label{curr22} \pa_{\mu} j^{(2)\,\mu}&=&0,\er where the currents
$j^{(n)}_{\mu} \,(n=1,2)$ are of the type given in
(\ref{current1})-(\ref{current2}). The conservation laws
(\ref{curr11})-(\ref{curr22}) can be verified as the result of the
field equations of motion \cite{jhep2}. As in the free
bi-fundamental fermion case the global $U(1)$ charge is associated
to the transformation $u_{2}u_{1}^{-1}$ and the currents
$j^{(1)}_{\mu}$\, and $j^{(2)}_{\mu}$ share the same charge. In
fact, the two currents are related by $\psi_{\a} \leftrightarrow
\psi_{\a}^{\dagger}$. Therefore the Noether $U(1)$ charge is given
by
 \br \label{charge} Q \equiv e_{n}\,\int d^2 x\, j^{(n)\, 0},\,\,\,\,n=1\,\, \mbox{or}\,\, 2;\er
where the $e_{n}$ are some normalization factors, which can be set
$e_{1}=1,\,e_{2}=-1$.

\subsection{The Thirring type interactions and NC bosonization }
\label{mthint}

 Let us bosonize the four-fermion interaction terms
of the model (\ref{ncmt0}), i.e. \br {\cal L}_{\l_{bf}}&=&
-\frac{\l_{bf}}{4} j^{(1)\,\mu} \star j^{(1)}_{\mu} -
\frac{\l_{bf}}{4} j^{(2)\,\mu}\star j^{(2)}_{\mu}
 \label{ncth1}.\er

Here we shall use directly the bosonization dictionary for the
currents established in (\ref{bosoni1})-(\ref{bosoni2})
\footnote{There exists another prescription to bosonize the
current-current coupling using the ``completing the square" type
Hubbard-Stratonovich identity which also holds in the NC case
\cite{moreno2}.}. Then, the four fermion interactions of
(\ref{ncth1}) correspond to the following terms in the boson
theory \br\label{currboso1} - \frac{\l_{bf}}{4} \int d^2x\,
2(\bar{\psi}\star \g_{+}\psi)\star (\bar{\psi}\star \g_{-}\psi)
\rightarrow \frac{\l_{bf}}{32\pi^2} \int d^2x\,\mbox{tr}\,
g_{1}^{-1}\star\pa_{+}g_{1}\star g_{1} \star
\pa_{-}g_{1}^{-1}\\
\label{currboso2} - \frac{\l_{bf}}{4} \int d^2x\, 2(\psi_{\b}\star
(\g_{+})_{\a\b}\bar{\psi}_{\a})\star (\psi_{\s}\star
(\g_{-})_{\d\s}\bar{\psi}_{\d}) \rightarrow
\frac{\l_{bf}}{32\pi^2} \int d^2x\,\mbox{tr}\,
g_{2}^{-1}\star\pa_{+}g_{2}\star g_{2} \star \pa_{-}g_{2}^{-1}.\er

It is known that the bosonized four fermion interaction term in
the commutative case adds to the kinetic term an extra
contribution, whereas the right hand side
 terms in (\ref{currboso1})-(\ref{currboso2}) contribute an
 infinite series of theta dependent higher derivative terms to the
   first quadratic term. In this way the NC extension provides
  additional terms making the bosonized theory in fact very different
  from the commutative version.

 The bosonized Lagrangian for the kinetic and four-fermion
 interaction terms becomes
\br\nonumber W[g_{1}] + W[g_{2}] +\frac{\l_{bf}}{32\pi^2} \int
d^2x\, \mbox{tr}\,g_{2}^{-1}\star\pa_{+}g_{2}\star g_{2} \star
\pa_{-}g_{2}^{-1}+\\\frac{\l_{bf}}{32\pi^2} \int d^2x\,\mbox{tr}\,
g_{1}^{-1}\star\pa_{+}g_{1}\star g_{1} \star
\pa_{-}g_{1}^{-1}\label{boso1}\er

In order to understand better the bosonic sector we need to
uncover a canonically normalized scalar field related to the
global $U(1)$ symmetry which corresponds to the charge of the
fermion field. So, let us consider the scalar fields $\L_{i}$
defined in (\ref{u1u1}). Then, the $\theta$ expansion of
(\ref{boso1}) provides \br \frac{1}{4\pi} \sum_{i=1}^{2}\int d^2x
\, \pa_{+}\L_{i}\pa_{-}\L_{i} + \frac{\l_{bf}}{32
\pi^2}\sum_{i=1}^{2}\int d^2x \,
\pa_{+}\L_{i}\pa_{-}\L_{i}+\,\,...\er

Taking into account these contributions and the definitions
(\ref{pm1}) one gets \br\frac{1}{8\pi} \(1+ \frac{\l_{bf}}{8\pi}\)
\int d^2x \, \[
\pa_{+}\L_{+}\pa_{-}\L_{+}+\pa_{+}\L_{-}\pa_{-}\L_{-}\] +\,\, ...,
\label{kinetic12}\er therefore, the canonically normalized scalar
related to the fermion charge corresponds to \br \Phi_{+}\equiv
\[\frac{1}{8\pi} \(1+ \frac{\l_{bf}}{8\pi}\)\]^{1/2} \L_{+}.
\label{canonic}\er

This result will be used below to relate the two-field NC
sine-Gordon and NCMT$_{1}$ coupling constants. Observe that for
stability of the bosonic theory we must have $\l_{bf} > -8\pi$.
The allowed range of the coupling gets enlarged as compared to the
anti-fundamental massless Thirring which is $\l_{f} > -4\pi$\,
\cite{nunez}. Recall that for the ordinary massless Thirring model
this bound is $\l > -\pi$ \cite{coleman}. Thus, there is a whole
range of coupling constants, depending on the fermion
representation, for which the NC massless Thirring is a sensible
model.

At this stage we have reproduced two copies of the WZNW action
 associated, respectively, to the fields $\L_{1}$ and $\L_{2}$, and
additional terms in (\ref{boso1}) due to the current-current
couplings. A two-field potential will emerge from the
bi-fundamental fermion mass term, as we will see in the next
subsection.

\subsection{The mass term and NC bosonization}
\label{mthmass}

Next we consider the bosonization of the bi-fundamental fermion
mass term. The relevant term is \br \label{mass} S_{M}= m\int d^2x
\(\psi^{\dagger\, j_{1}}_{R\,\,i_{1}}\star
\psi_{L\,\,j_{1}}^{i_{1}} +\psi^{\dagger\,
j_{1}}_{L\,\,\,i_{1}}\star \psi_{R\,\,j_{1}}^{i_{1}}\) \er

Here, we follow the procedure implemented in Refs. \cite{Burgess1,
nunez} in which the chiral symmetry plays a central role in the
bosonization process. In our case this procedure still applies
since the free bi-fundamental fermion possesses a global axial
symmetry $U_{5}(1)\times U_{5}(1)$ for which the starred products
collapse to the ordinary ones. In fact, the {\sl free} fermion
theory (\ref{partifree1}) is symmetric under the global chiral
symmetry \br \label{u1u1qui} (\psi)^{T} \rightarrow
\(e^{i\a_{2}\g_{5}} \psi \)^{T} e^{i\a_{1}\g_{5}}
;\,\,\,\,(\bar{\psi})^{T} \rightarrow \(e^{i\a_{1}\g_{5}}
\bar{\psi}\)^{T} e^{i\a_{2}\g_{5}},\er where $T$ stands for matrix
transpose operation and the suggestive products have been written
in spite of their global character. It is assumed that the bosonic
theory preserves this symmetry \cite{Witten1}. Besides, the
duality approach to bosonization relies on gauging the vector
$U(1)\times U(1)$ symmetries \cite{Burgess1, Burgess2}. On the
other hand, it is a known fact that the axial symmetry does not
survive quantization. The relevant axial anomalies in NC plane are
\cite{moreno1, moreno2, nakajima, ardalan} \br\label{axial1}
\pa_{\mu} j_{5}^{(2)\mu} &=& \frac{1}{2\pi}\, \epsilon^{\mu\nu}
F_{\mu\nu}^{(2)}, \,\,\,\,\,\,\,\,\,\,\,        j_{5}^{(2)\mu}=
\psi_{\b} \star
(\g^{\mu}\g_{5})_{\a\b}\bar{\psi}_{\a}   , \\
\pa_{\mu} j_{5}^{(1)\mu} &=& -\frac{1}{2\pi}\, \epsilon^{\mu\nu}
F_{\mu\nu}^{(1)},\,\,\,\,\,\,\,\,\,j_{5}^{(1)\mu}=\bar{\psi} \star
\g^{\mu}\g_{5} \psi,\label{axial2},\er where the
$F_{\mu\nu}^{(i)}$ are the NC gauge field strengths.

In order to make the axial symmetry manifest during the
bosonization process the authors of \cite{Burgess1} provided the
lagrange multiplier with a transformation property to cancel the
 fermion axial anomaly. In fact, the gauged action would have the
 terms
\br \sim \mbox{exp}\( \int \-d^2x \[\, \frac{1}{2\pi} \L_{1}\star
\epsilon^{\mu\nu} F_{\mu\nu}^{(1)} - \frac{1}{2\pi} \L_{2}\star
\epsilon^{\mu\nu} F_{\mu\nu}^{(2)}\] \) \label{multipliers}\er

The key point is that if the fields $\L_{i}$ transform as $\L_{i}
\rightarrow \L_{i}-\a_{i}$ under axial transformations, then the
Lagrange multiplier terms (\ref{multipliers}) will cancel the
axial anomaly terms (\ref{axial1})-(\ref{axial2}) which emerge in
the path-integral, thus making the whole generating functional
manifestly chiral symmetric. These considerations tell us how the
$\L_{i}$ fields transform under chiral rotations \cite{Burgess1}.

Even though the mass term is no longer axial symmetric, these
transformation properties allow us to bosonize the mass term. The
observation is that these transformation rules  remain the same
for the bilinear terms in (\ref{mass}). The problem resides in
finding the bosonic functionals associated to these bilinear
terms,  ${\cal F}(\L_{1},\, \L_{2}) \equiv \psi^{\dagger\,
j_{1}}_{R\,\,i_{1}}\star \psi_{L\,j_{1}}^{i_{1}}$ and its
corresponding ${\cal F}^{\dagger}(\L_{1},\, \L_{2})$. Here the
functionals depend on $\L_{i}$ through $\star$-products. Under the
{\sl global} symmetry transformations (\ref{u1u1qui}) one has \br
\psi^{\dagger\, j_{1}}_{R\,\,i_{1}}\star \psi_{L\,j_{1}}^{i_{1}}
\rightarrow e^{-2i(\a_{1}+ \a_{2})} \psi^{\dagger\,
i_{1}}_{R\,j_{1}}\star \psi_{L\,j_{1}}^{i_{1}},\er and taking into
account the $\L_{i}$ field transformations discussed above we have
the functional transformation property \br\label{funceq} {\cal
F}(\L_{1}-\a_{1},\, \L_{2}-\a_{2}) = e^{-2i(\a_{1}+ \a_{2})} {\cal
F}(\L_{1},\, \L_{2}).\er

Therefore, one has that \br\label{solution} {\cal F}(\L_{1},\,
\L_{2})\sim e^{2i\L_{1}}_{\star} \star e^{2i\L_{2}}_{\star},\er
uniquely solves  (\ref{funceq}) for globally defined rotations.
With this result the mass term (\ref{mass}) becomes
 \br \label{potential} S_{M}= \int d^2x \, m \a_{0}\(e^{2i\L_{1}}_{\star} \star
e^{2i\L_{2}}_{\star}+e^{-2i\L_{2}}_{\star} \star
e^{-2i\L_{1}}_{\star} \),\er where $\a_{0}$ is a parameter to be
associated to the zero point energy.

This is precisely the potential proposed by Lechtenfel et al. for
the NC sine-Gordon \cite{lechtenfeld}. In the previous section we
associated the field $\L_{+}=(\L_{1}+\L_{2})$ to the charged
sector of the bi-fundamental NC free fermion model. It is expected
that the same field will be associated to the global $U(1)$ charge
of the NCMT$_{1}$ theory as discussed in the paragraph after the
Eqs. (\ref{curr11})-(\ref{curr22}). This field has been isolated
from the quadratic and quartic terms in (\ref{kinetic12}). So, we
would like to uncover this field in the bosonized mass term
(\ref{potential}) as well. Actually, this is not possible to do
exactly due to the $\star-$product involved and only one can
pursue it in the $\theta$ perturbation expansion. Then the
expansion of the potential provides \br \int d^2x \, m \a_{0}\,
\[2\,\mbox{cos}(2\L_{+})+\,\,...\].\er

The canonically normalized field $\Phi_{+}$ was defined in
(\ref{canonic}), then the first potential term becomes \br \int
d^2x \, m \a_{0}\,
\[2\,\mbox{cos}(\b_{bf}\, \Phi_{+})+\,\,...\],\er
where \br\label{duality} \frac{32\pi}{\b^2_{bf}}= 1+
\frac{\l_{bf}}{8\pi}.\er

Thus, we have a S-duality relationship. The relation
(\ref{duality}) constitutes the quantum version of the strong-weak
duality obtained in the classical $1-$soliton sector of the
NCMT$_{1}$ $\rightarrow$ NCSG$_{1}$ correspondence \cite{jhep2}.

This type of relationship has also been derived in  \cite{nunez}
 for the NC bosonization of the massive Thirring model with
  anti-fundamental fermion (see eq. (\ref{dual01})). With these results in hand one can establish that the couplings of
  the various models are related by
 \br
 \label{relationships}
 \l_{bf}= 2\, \l_{\bar{f}}=
8\, \l_{MT},\,\,\,\,\,\,\,\b^2_{bf}= 2\, \b^2_{\bar{f}}= 8\,
\b^2_{SG},
 \er
where the $\l_{\bar{f}}$, $\b^2_{\bar{f}}$ couplings correspond to
the fermionic and bosonic sector considered in \cite{nunez},
respectively \footnote{It is expected that by formally setting
$\l_{f}\equiv \l_{\bar{f}},\,\b^2_{f}\equiv \b^2_{\bar{f}}$ in
(\ref{dual01}) one can obtain the same duality relationship for
fundamental fermion.}. The couplings $\l_{MT},\, \b^2_{SG}$ were
defined in (\ref{mt0})-(\ref{sg0}) for the models in ordinary
spacetime.

Notice that the duality relation is maintained in each term of the
$\theta$ expansions of the potential terms (\ref{potential}), as
well as the quadratic and quartic terms of (\ref{boso1}). To see
this fact notice that in order to define the canonically
normalized field $\Phi_{+}$ the fields $\L_{i}$ have been
re-scaled as $\L_{i} \rightarrow \frac{\beta}{2} \L_{i}$, then the
expansions will only produce terms proportional to positive powers
of the bosonic coupling constant $\b$.

In the bosonic sector we must have a topological charge
corresponding to the global $U(1)$ Noether charge of the fermionic
 theory. The physical scalar field associated to this topological
charge may be identified in the commutative limit $\theta
\rightarrow 0$ of the model. The field $\L_{-}$ decouples
 completely in this limit since the quartic terms contributions are only of the bilinear
 kinetic type and the combination
 $\L_{+}$ alone appears in the potential term. So, it is clear that the field
$\Phi_{+}$ defined in (\ref{canonic}) carries this charge in the
bosonic sector. Then, we can define \cite{grisaru1,
jhep2}\footnote{The bosonization rules
(\ref{bosoni1})-(\ref{bosoni2}) and the Noether charge definition
(\ref{charge}) might suggest for $j_{top}^{\mu}$ to be some
combinations of the type $e^{\pm i\L_{j}}_{\star} \star \pa_{\pm}
e^{\mp i\L_{j}}_{\star}, \,(j=1,2)$. However, it is a fact that
only the field $\Phi_{+}(\sim \L_{+})$ appears in the commutative
limit of the potential and since the $Q_{top}$ is related to a
discrete symmetry of the potential it is plausible to define it as
in (\ref{topo}).}

\br j^{\mu}_{top}\equiv \frac{\b_{bf}}{2\pi}
\epsilon^{\mu\nu}\pa_{\nu}
\Phi_{+},\,\,\,\,\,\,\,\,Q_{top}=\frac{\b_{bf}}{2\pi}
\int_{-\infty}^{\infty} dx^{1} \frac{d \Phi_{+}}{dx^1}\equiv
\sum_{n=0} \theta^{n} Q_{n},\label{topo}\er where $Q_{top}$ is the
topological charge expressed as a power series in $\theta$. This
expression is the same as in the classical NCSG$_{1}$ model. In
fact, for the  classical one (anti-)soliton sector it has been
obtained an exact topological charge equal to $\pm 1$,
respectively; i.e. one has in (\ref{topo}), $Q_{n}=0 \, (n>0)$,
since the $1-$soliton solution of the ordinary sine-Gordon model
solves also the NC theory \cite{lechtenfeld, jhep2, dimakis}. The
Noether and topological currents equivalence is also present in
the ordinary MT/SG relationship, on the quantum \cite{coleman,
Naon} and classical level establishing a soliton/particle
correspondence \cite{nucl2, orfanidis}.

Finally, the NCMT$_{1}$ model (\ref{ncmt0}) is equivalent to a
bosonic model which comprises the eq. (\ref{boso1}) plus the eq.
(\ref{potential}), which we record below in matrix notation
\br\nonumber {\cal S}_{bf}[g_{1}, g_{2}] &=& W[g_{1}] + W[g_{2}] +
\int d^2x\,\mbox{tr} \[ m \a_{0}\, \(\s_{1} g_{2}^{-2}\star
g_{1}^{-2}\s_{1} \star g_{1}^{2} \star g_{2}^{2}\) +\\&&
 \frac{\l_{bf}}{32\pi^2} \, \(
g_{2}^{-1}\star\pa_{+}g_{2}\star g_{2} \star \pa_{-}g_{2}^{-1}+
g_{1}^{-1}\star\pa_{+}g_{1}\star g_{1} \star \pa_{-}g_{1}^{-1} \)
\] \label{ncsg12}, \er
where the matrices $g_{1}, g_{2}$ are given in (\ref{u1u1}) and
$\s_{1} = [\begin{array}{cc}
0 &  1 \\
1 & 0
\end{array}].$ Thus, the bosonic sector comprises two WZW actions plus a
potential depending on two scalar fields (thus, the first three
terms of (\ref{ncsg12}) correspond to the Lechtenfeld et al.
proposal \cite{lechtenfeld} for NC sine-Gordon) and additional
quartic terms coming from the current-current interactions. These
quartic terms were also absent in the NCSG$_{1}$ model in the
purely classical treatments of \cite{jhep2}.

The upper and lower indices of a model defined for bi-fundamental
matter can be interchanged since the metric is $\d_{AB}$ in the
normalization $\mbox{tr}(T_{A}T_{B})=\frac{1}{2} \d_{AB}$ of the
Lie algebra generators. This symmetry can be written manifestly
using the group index notation as in (\ref{free0}) or (\ref{mass})
for the NCMT$_{1}$ model (\ref{ncmt0}). Notice that the symmetry
$i \leftrightarrow j$ which interchanges the two fundamental
representations corresponds to the symmetry $g_{1} \leftrightarrow
g_{2}$ of the bosonic sector
 (\ref{ncsg12}) (recall that $g_{1}\star g_{2}=g_{2}\star g_{1}$ ). The action (\ref{ncsg12}) can be written in a
 more compact form using the Polyakov-Wiegmann identity and the
 representation (\ref{u1u1}) for $g\in U(1)\times U(1)$
 \br {\cal S}_{bf}[g] &=& W[g] +\int d^2x\,\mbox{tr} \[ m \a_{0}\, \(\s_{1} g^{-2}
\s_{1}\star g^{2}\) + \frac{\l_{bf}}{32\pi^2} \, \(
g^{-1}\star\pa_{+}g\star g \star \pa_{-}g^{-1}\)
\] \label{ncsg121}. \er

This representation is more convenient in order to compare it to
its analog model obtained from reduction of the parent action, the
$GL(2, C)$ NC affine Toda model coupled to matter (Dirac) fields
\cite{jhep2}. Except for the last quartic term, this is the
NCSG$_{1}$ model defined in \cite{jhep2, lechtenfeld}. Recall that
the theories NCSG$_{1, 2}$ defined in \cite{jhep2} correspond to
the Lechtenfeld et al. and Grisaru-Penati models, respectively.
The Grisaru-Penati model has the potential $(e^{i\phi}_{\star}+
e^{-i\phi}_{\star})$ corresponding to one complex scalar field,
instead of the potential (\ref{potential}) defined for two fields
of the Lechtenfeld et al. model. Strictly speaking, in our
constructions above the fields $g_{i} \in U(1)_{C}$ (complexified
$U(1)$), because the determinants were computed in Euclidean
space. This is the case in ordinary commutative space, in which
the complexification of $U(1)$, $U(1)_{C}$ is considered
\cite{naculich}. In the construction of the model NCSG$_{1}$ we
may consider the abelian subgroup of $GL(2,C)$, $U(1)_{C}\times
U(1)_{C}$ to which the fields $g_{1}$ and $g_{1}$ belong,
respectively \cite{jhep2}. Nevertheless, it is a fact that the
NCSG$_{1}$ model possesses real soliton type solutions for the
field $\L_{+}(= \L_{1}+\L_{2})$, whereas $\L_{-}(= \L_{1}-\L_{2})$
remains as a decoupled free field \cite{jhep2, lechtenfeld}.

\section{Conclusions and discussions}
\label{conclusion}

The NC bosonization of the NCMT$_{1}$ model and some  properties
emerging from this process have been considered. In order to
bosonize the current-current interactions we have used the
bosonization dictionary (\ref{bosonimatrix})
 established for the relevant currents of a  bi-fundamental free
 fermion. In order to bosonize the mass term we have used the
 axial transformation rules that the fields $\L_{i}$ must undergo in order to
 maintain these symmetries manifest during the bosonization process of a free fermion as developed
 in NC space \cite{nunez}, following
 the ordinary space approach of
 \cite{Burgess1}. Even though the mass term breaks the both axial symmetries
 the axial transformation rules remain the same, thus allowing to construct
 the boson sector of the fermion mass term. The bosonized mass
 term reproduces the Lechtenfeld et al. proposal for the NC sine-Gordon
 potential \cite{lechtenfeld, jhep2}.
  It is uncovered the duality relationship between the coupling
constants of the both sectors (\ref{duality}), as well as certain
relationships between the couplings for the models with
bi-fundamental, (anti-)fundamental fermion representations and
their bosonic sectors and the ordinary MT/SG
 couplings (\ref{relationships}). The global $U(1)$ charge of the
 fermionic sector (\ref{charge}) defined for either, the current $j^{(1)\,\mu}$ or
 $j^{(2)\,\mu}$, corresponds to the topological charge (\ref{topo}) defined for the
 field $\Phi_{+}(\sim \L_{+})$. It is shown that the auxiliary field $\L_{-}$
 decouples in the commutative limit. However the deformed
 situation requires the both scalar fields $\L_{\pm}$.

Disregarding for the moment the quantum nature of the terms, such
as certain normal ordering prescriptions, the bosonic
(\ref{ncsg12})[or (\ref{ncsg121})] and fermionic (\ref{ncmt0})
models resemble to their classical counterparts NCSG$_{1}$ and
NCMT$_{1}$, respectively,  as defined in \cite{jhep2}(except for
the last quartic term in (\ref{ncsg121}). A remarkable feature of
the NCSG$_{1}$ model, as well as some NC integrable systems
\cite{dimakis}, is that for $1-$(anti)soliton type solution the
$\star-$products in its equations of motion collapse to ordinary
ones since one has in general $f(x-v t) \star g(x-v t)= f(x-v t)
g(x-v t)$. Then the $\star-$products for these type of fields in
the equations of motion of the model (\ref{ncsg121}) reduce to the
ordinary ones. In particular, the contributions from the last
quartic term in (\ref{ncsg121}) are only of the type emerging from
the kinetic terms of the action. So, we may say that regarding the
classical $1-$(anti)soliton type solution, the bosonized model
(\ref{ncsg121}) and the classical NCSG$_{1}$ theories are very
similar. This is in contrast to the usual MT/SG duality in which
the corresponding classical and quantum Lagrangians look very
similar, apart from the field renormalizations and the relevant
quantum corrections to the coupling constants.

Various aspects of the bosonized  model derived above  deserve
attention in future research, e.g. the properties of the S-matrix,
the NC zero-curvature formulation and integrability properties,
the NC multi-soliton spectrum and their scattering properties. The
ordinary equivalence SG/MT has been used to understand some
confinement mechanism in QCD$_{2}$ with one flavor and N colors
\cite{nucl1}, on this regard it would be interesting to consider
the NC situation. Moreover, the multi-fermion extensions of the
NCMT$_{1}$ type model also deserve attention, e.g. in connection
to multi-flavor noncommutative QCD$_{2}$. In ordinary space the
correspondence between the multi-fermion massive Thirring and
multi-boson sine-Gordon models related to any (untwisted) affine
Lie algebra has been considered in \cite{jhep1} on the classical
level. The classical correspondence between three-fermion massive
Thirring and three-boson sine-Gordon has been addressed in
\cite{jmp} and the quantum field theory aspects studied in
\cite{epjc} through bosonization techniques. It would be
interesting to address their NC extensions.

\vskip 0.4in

                {\sl Acknowledgments}

\vskip 0.2in The author thanks H.L. Carrion and M. Rojas for
collaboration in a previous work and to ICET and Prof. M. C.
Ara\'ujo at the Mathematics Department-UFMT
                for hospitality. This work has been supported by CNPq-FAPEMAT.


\begin{thebibliography}{**}
\bibitem{seiberg}
N.~Seiberg and E.~Witten,
\JHEP{9909}{1999}{032};\\
A.~Connes, M.~R.~Douglas and A.~Schwarz,
\JHEP{9802}{1998}003;\\
M.~R.~Douglas and C.~M.~Hull,
\JHEP{9802}{1998}008.
\bibitem{Lechtenfeld1}
  O.~Lechtenfeld, A.~D.~Popov and B.~Spendig,
\PLB{507}{2001}{317}.
\bibitem{Takasaki}
  K.~Takasaki,
  \JGP{37}{2001}{291}.
  \bibitem{lechtenfeld}
 O. Lechtenfeld, L. Mazzanti, S. Penati, A. D. Popov, L. Tamassia,
 \NPB{705}{2005}{477}.
                \bibitem{grisaru1}
                M. T. Grisaru and S. Penati, \NPB{655}{2003}{250}.
\bibitem{grisaru2}
                M.T. Grisaru, L. Mazzanti, S. Penati and
                L.Tamassia, \JHEP{0404}{2004}{057}.
                \bibitem{cabrera}
                I.~Cabrera-Carnero and M.~Moriconi,
                \NPB{673}{2003}{437}.
\bibitem{jhep2} H. Blas, H.L. Carrion and M. Rojas, \JHEP{0503}{2005}{037}; hep-th/0502051.
 \bibitem{coleman}
                S. Coleman, \PRD{11}{1975}{2088};\\
                S. Mandelstam, \PRD{11}{1975}{3026}.
 \bibitem{nucl2}
                H. Blas, \NPB{596}{2001}{471};\\
                H. Blas and B.M. Pimentel, \AoP{282}{2000}{67}.
                \bibitem{matter}
L.A.~Ferreira, J.L.~Gervais, J.~Sanchez~Guillen and M.V.~Savelev,
\NPB{470}{1996}{236}.
                \bibitem{jhep1}
                H. Blas, \JHEP{0311}{2003}{054}.
         \bibitem{nunez} C.~Nunez,
K.~Olsen and R.~Schiappa, \JHEP{0007}{2000}{030}.
\bibitem{moreno1} E.~F.~Moreno and
F.~A.~Schaposnik, \JHEP{0003}{2000}{032}.
\bibitem{moreno2}
E.~F.~Moreno and F.~A.~Schaposnik,
\NPB{596}{2001}{439}.
                \bibitem{Naon}
                C. M. Naon,
\PRD{31}{1985}{2035}.
\bibitem{Burgess1} C.~P.~Burgess and
F.~Quevedo,
\NPB{421}{1994}{373}.
\bibitem{Burgess2}
C.~P.~Burgess and F.~Quevedo,
\PLB{329}{1994}{457}.
\bibitem{Guillou}
J.~C.~Le Guillou, E.~Moreno, C.~Nunez and F.~A.~Schaposnik,
\NPB{484}{1997}{682}.
\bibitem{Witten1} E. Witten,
\CMP{92}{1984}{455}.
\bibitem{Ganor} O. J. Ganor, G. Rajesh and S.
Sethi,
\PRD{62}{2000}{125008};\\R. Gopakumar, J. M. Maldacena, S.
Minwalla and A. Strominger,
\JHEP{0006}{2000}{036};\\
  J.~X.~Lu, S.~Roy and H.~Singh,
  \JHEP{0009}{2000}{020};\\
  R.~G.~Cai and N.~Ohta,
  \PTP{104}{2000}{1073}; \JHEP{0003}{2000}{009}; \\
R.G. Cai, J.-X. Lu and N. Ohta, \PLB{551}{2003}{178}.
\bibitem{Gomis}
  N.~Seiberg, L.~Susskind and N.~Toumbas,
  \JHEP{0006}{2000}{044}.\\
J. Gomis, T. Mehen, \NPB{591}{2000}{265}.
\bibitem{Chu1}
C-S. Chu, J. Lukierski, W.J. Zakrzewski, \NPB{632}{2002}{219}.
\bibitem{Furuta}
K. Furuta and T. Inami, \MPLA{15}{2000}{997}.
\bibitem{gracia-bondia}
J.~M.~Gracia-Bondia and C.~P.~Martin,
\PLB{479}{2000}{321}.
\bibitem{liao} Y.~Liao and K.~Sibold,
\PLB{586}{2004}{420}.
\bibitem{terashima}
S. Terashima, \PLB{482}{2000}{276}.
\bibitem{nakajima} T.~Nakajima,
\PRD{68}{2003}065014.
\bibitem{Bonora} L. Bonora and A. Sorin,
\PLB{521}{2001}{421}.
\bibitem{Martin}
  C.~P.~Martin,
  \NPB{623}{2002}{150}.
\bibitem{ardalan} F.~Ardalan and N.~Sadooghi,
\IJMPA{16}{2001}{3151};
\IJMPA{17}{2002}{123}.
\bibitem{dimakis}
A.~Dimakis and F.~Mueller-Hoissen, {\sl A noncommutative version
of
  the nonlinear schroedinger equation}, {\sl hep-th}/0007015.
\bibitem{orfanidis}
S.~J.~Orfanidis,
  \PRD{14}{1976}{472};\\ S.~J.~Orfanidis and R.~Wang,
  \PLB{57}{1975}{281}.
                \bibitem{naculich} S.~G.~Naculich and
H.~J.~Schnitzer,
\NPB{332}{1990}{583};\\ R.~E.~Gamboa Saravi, F.~A.~Schaposnik and
J.~E.~Solomin,
  \PRD{30}{1984}{1353}.
\bibitem{nucl1}
                H. Blas and L.A. Ferreira, \NPB{571}{2000}{607};\\
                H. Blas, \PRD{66}{2002}{127701}.
\bibitem{jmp}J. Acosta, H. Blas, \JMP{43}{2002}{1916}, see also {\sl hep-th}/0407020;
\bibitem{epjc}
              H. Blas, \EPJC{37}{2004}{251}, see also {\sl hep-th}/0409269.
\end{thebibliography}
\end{document}